\documentclass[a4paper,10pt]{article}
\usepackage[dvips]{graphicx}
\usepackage{amssymb,amsmath}
\numberwithin{equation}{section}

\begin{document}

\title{Reconstruction of   f-essence and fermionic Chaplygin gas models of dark energy}
\author{P.Tsyba, K.Yerzhanov, K.Esmakhanova,   I.Kulnazarov,  G.Nugmanova,  R.Myrzakulov\footnote{The corresponding author. Email: rmyrzakulov@csufresno.edu; rmyrzakulov@gmail.com}\\ \textit{Eurasian International Center for Theoretical Physics, Dep. Gen. $\&$ Theor.  } \\ \textit{ Phys., Eurasian National University, Astana 010008, Kazakhstan} }

\date{}

\maketitle
\begin{abstract} 

Recently, it was proposed a new fermionic model of dark energy, the so-called f-essence. In this work,  we first present fermionic Chaplygin gas models of dark energy corresponding to the usual Einstein-Dirac theory. Then we  explicitly reconstruct the different f-essence models. In particular, these models  include the fermionic  Chaplygin gas and the fermionic generalized Chaplygin gas models of dark energy.  We also derive the equation of state parameter of the selected f-essence models. 
\end{abstract}
\vspace{2cm} 

\sloppy

\tableofcontents
\section{Introduction} 
The late time accelerated expansion of the universe has attracted much attention in the recent years. Direct evidence of cosmic acceleration comes from high red-shift supernova experiments \cite{Perlmutter}-\cite{Riess}. Some other observations, such as cosmic microwave background fluctations  and large scale structure, provide an indirect evidence. These observations seem to chage the entire picture of our matter filled universe. It is now believed that most part of the universe contains dark matter and dark energy. The extensions and/or modifications of general relativity  seem attractive to explain late time acceleration and dark energy. Friedmann-Robertson-Walker (FRW) models, being spatially homogeneous and isotropic in nature, are best for the representation of the large scale structure of the present universe.  Several theoretical models of dark energy were proposed. Among of dark energy 
 models, interesting attempts are based on the  k-essence, a scalar field with noncanonical kinetic terms which was first introduced as a model for inflation \cite{Mukhanov1}-\cite{Mukhanov3} (see also \cite{Linder}-\cite{Vikman}). Later, several fermionic dark energy models were introduced \cite{Ribas1}-\cite{Boehmer3}. 
 Recently, new models of dark energy, the so-called f-essence and 
 g-essence were proposed   \cite{MR1}.

Note that the modern constraints on the equation of state (EoS) parameter $\omega$ of dark energy are around the cosmological constant value ($\omega=-1$) and a possibility that $\omega$ is varied in time is not excluded. From the theoretical point of view there are several essentially different cases:
 \begin{eqnarray}
	\omega&<&-1, \, \, \, \,\,\,\,\,\,\,\,\,\,\,\,\,\,\, \, \, \,\,\,\,\,\,\,\,\,\,\,\,\,\, \, \, \, \,\,\,\,\,\,\,\,\,\,\,\,\,\,(phantom \quad matter),\\ 
			\omega&=&-1, \, \, \, \,\,\,\,\,\,\,\,\,\,\,\,\,\, \, \, \, \,\,\,\,\,\,\,\,\,\,\,\,\,\,\, \, \, \,\,\,\,\,\,\,\,\,\,\,\,\,\, (cosmological \quad constant),\\
			\omega&\in&(-1, -1/3), \, \, \, \,\,\,\,\,\,\,\,\,\,\,\,\,\,\, \, \, \,\,\,\,\,\,\,\,\,\,\, (quintessence),\\ 
			\omega&=&0, \, \, \, \,\,\,\,\,\,\,\,\,\,\,\,\,\, \, \, \, \,\,\,\,\,\,\,\,\,\,\,\,\,\,\, \, \, \,\,\,\,\,\,\,\,\,\,\,\,\,\,\,\,\,\,\,\,(dust),\\ 
			\omega&=&1/3, \, \, \, \,\,\,\,\,\,\,\,\,\,\,\,\,\,\, \, \, \,\,\,\,\,\,\,\,\,\,\,\,\,\, \, \, \, \,\,\,\,\,\,\,\,\,\,\,\,\,\,(radiation),\\ 
			\omega&\in&(1/3, 1), \, \, \, \,\,\,\,\,\,\,\,\,\,\,\,\,\, \, \, \, \,\,\,\,\,\,\,\,\,\,\,\,\,\, \, \, \, \,\,\,(hard  \quad 
			Universe),\\ 
\omega&=&1, \, \, \, \,\,\,\,\,\,\,\,\,\,\,\,\,\, \, \, \, \,\,\,\,\,\,\,\,\,\,\,\,\,\,\, \, \, \,\,\,\,\,\,\,\,\,\,\,\,\,\, \,\,\,\,\, (stiff  \quad matter),\\ 
\omega&>&1, \, \, \, \,\,\,\,\,\,\,\,\,\,\,\,\,\, \, \, \, \,\,\,\,\,\,\,\,\,\,\,\,\,\,\, \, \, \,\,\,\,\,\,\,\,\,\,\,\,\,\, \,\,\,\,\, (ekpyrotic \quad matter).
	\end{eqnarray} 
	It is also important to find a model which follows from the fundamental principles and describes a crossing of the $\omega=-1$ barrier.
	
In this paper, we focuss our attention to explore the explicit models 
 of f-essence. Using the reconstruction method (see e.g. \cite{Odin}-\cite{MR2}), we explicitly construct some particular reductions of f-essence in the FRW universe. For example, we find the fermionic Chaplygin gas and the fermionic generalized Chaplygin gas models of dark energy.
 The corresponding equation of state parameters are presented. 

 \section{Some basics of f-essence}

The action of   f-essence reads as  
\cite{MR1}
\begin {equation}
S=\int d^{4}x\sqrt{-g}[R+2K( Y,  \psi, \bar{\psi})],
\end{equation} 
 where $R$ denotes the scalar curvature (the Ricci scalar). Here  $L_f=2K( Y,  \psi, \bar{\psi})$ is the Lagrangian density of the fermionic field, where $\psi=(\psi_1, \psi_2, \psi_3, \psi_4)^{T}$  is a fermionic function  and $\bar{\psi}=\psi^{\dagger}\gamma^0$ is its adjoint function, the dagger represents complex conjugation. The canonical kinetic term for the fermionic field is
\begin {equation}
Y=0.5i[\bar{\psi}\Gamma^{\mu}D_{\mu}\psi-(D_{\mu}\bar{\psi})\Gamma^{\mu}\psi],
\end{equation}
where   $ D_{\mu}$ is a covariant derivative. Note that the fermionic fields are treated here as classically commuting fields.  

In general, the equations corresponding to the action (2.1) have very complicated form. We are here modest and consider the simple cosmological metric, namely,  the  homogeneous, isotropic and flat FRW universe filled with f-essence. This metric is given by 
	\begin{equation}
ds^2=dt^2-a^2(dx^2+dy^2+dz^2)
\end{equation}
and the vierbein is chosen to be (see e.g. \cite{Armendariz-Picon})
	\begin{equation}
	(e_a^\mu)=diag(1,1/a,1/a,1/a),\quad 
(e^a_\mu)=diag(1,a,a,a).
\end{equation}

 The preliminary set-up for writing the equations of motion is now complete. So for  the FRW metric (2.3), the equations corresponding to the action (2.1) look  like \cite{MR1}
	\begin{eqnarray}
	3H^2-\rho&=&0,\\ 
		2\dot{H}+3H^2+p&=&0,\\
				K_{Y}\dot{\psi}+0.5(3HK_{Y}+\dot{K}_{Y})\psi-i\gamma^0K_{\bar{\psi}}&=&0,\\ 
K_{Y}\dot{\bar{\psi}}+0.5(3HK_{Y}+\dot{K}_{Y})\bar{\psi}+iK_{\psi}\gamma^{0}&=&0,\\
	\dot{\rho}+3H(\rho+p)&=&0,
	\end{eqnarray} 
where  a dot denotes a time derivative and $K_Y=dK/dY$. Here  the kinetic terms, the energy density  and  the pressure  take the form
\begin{equation}
 Y=0.5i(\bar{\psi}\gamma^{0}\dot{\psi}-\dot{\bar{\psi}}\gamma^{0}\psi)
  \end{equation}
  and
  \begin{equation}
  \rho=K_{Y}Y-K,\quad
p=K,
  \end{equation}
 respectively.  Also we remark that as $K=Y-V(\bar{\psi}, \psi)$, the model (2.1) gives the usual Einstein-Dirac theory. 
\section{Reconstruction of f-essence. Example: $K=F(Y)V(\bar{\psi}, \psi)$}
In this section,   we consider the following f-essence model 
\begin{equation}
  K=F(Y)V(\bar{\psi}, \psi).
  \end{equation}
  Then the energy density and pressure are given by
  \begin{equation}
  \rho=V(YF^{'}-F),\quad
p=VF,
  \end{equation}where $F^{'}=dF/dY$. From
  \begin{equation}
  YK_{Y}=-0.5(K_{\psi}\psi+K_{\bar{\psi}}\bar{\psi})  \end{equation}
  follows that
    \begin{equation}
  VYF^{'}=-0.5F(V_{\psi}\psi+V_{\bar{\psi}}\bar{\psi}).  \end{equation}
  If we assume that $V=V(u),\quad (u=\bar{\psi}\psi)$ then we obtain
   \begin{equation}
  VYF^{'}=-FV_{u}u  \end{equation}
  and 
   \begin{equation}
  u=\frac{c}{a^3VF^{'}},\quad c=const.  \end{equation}
 
  Now we consider a more simple case namely $V=mu^n$. Then we have
  \begin{equation}
  VYF^{'}=-FV_{u}u.  \end{equation}
  Hence we otain
   \begin{equation}
  F=\eta Y^{-n},\quad \eta=const.\end{equation}
  Similarly we have
   \begin{equation}
  u=(-\frac{c}{nm\eta a^3})^{\frac{1}{n+1}}Y.  \end{equation}
  Note that in this case the energy density and the pressure take the form
  \begin{equation}
  \rho=-m\eta(n+1) (\frac{u}{Y})^n,\quad
p=m\eta (\frac{u}{Y})^n.
  \end{equation}Hence the EoS parameter is
   \begin{equation}
  \omega=-\frac{1}{n+1}.
  \end{equation}So if $n<2$ this particular f-essence model describes the accelerated expansion of the universe. From 
   \begin{equation}
  -2\dot{H}=\rho+p=YK_Y=VYF^{'}=-nm\eta(\frac{u}{Y})^{n} \end{equation}
  we get
   \begin{equation}
  u=Y\sqrt[n]{\frac{2\dot{H}}{\eta nm}}.\end{equation}
  Finally  we come to the following equation for the scale factor   \begin{equation}
  2\dot{H}=\sqrt[n+1]{\eta nm(-c)^n}a^{-\frac{3n}{n+1}}. \end{equation}
  
   As an example, let us consider the power-law case : $a=a_0t^l$. Then from (3.14) we get
   \begin{equation}
  l=\frac{2(n+1)}{3n}, \quad a_0=[-\frac{3n}{4(n+1)}]^{\frac{n+1}{3n}}[\eta nm(-c)^n]^{\frac{1}{3n}}. \end{equation}
  The energy density and the pressure take the form
  \begin{equation}
  \rho=\frac{4(n+1)^2}{3n^2}t^{-2},\quad
p=-\frac{4(n+1)}{3n^2}t^{-2}.
  \end{equation}
  So we have 
  \begin{equation}
  u=Y\sqrt[n]{-\frac{4(n+1)}{3m\eta n^2t^{2}}}.
  \end{equation}
  The solution of the Dirac equation we can construct using the following formulas
  \begin{equation}
  \psi_j=\frac{c_j}{\sqrt{a^3K_Y}}e^{i\int (\gamma^0K_uK_Y^{-1})_jdt}\end{equation}
  but we drop it. Here $c=c_1^2+c_2^2-c_3^2-c_4^2$. 

Let us now we consider the case when the EoS parameter $\omega=\omega_0=const$. Then we have the following f-essence model
\begin{equation}
  K=V(\bar{\psi}, \psi)Y^{\frac{\omega_0+1}{\omega_0}}.
  \end{equation}
  For this case  we get
   \begin{equation}
  \rho=\frac{1}{\omega_0}V(\bar{\psi}, \psi)Y^{\frac{\omega_0+1}{\omega_0}},\quad
p=V(\bar{\psi}, \psi)Y^{\frac{\omega_0+1}{\omega_0}}
  \end{equation}so that 
  \begin{equation}
  p=\omega_0\rho.
  \end{equation}If $V=V(u)$ then for this example the potential takes the form
  \begin{equation}
  V=V_0u^{-\frac{\omega_0+1}{\omega_0}}.
  \end{equation}
  \section{Fermionic Chaplygin gas models of dark energy from the Einstein-Dirac equation}
  In the next two sections, we want to present the particular f-essence models which give the Chaplygin gas models of dark energy. In other words, accordingly to the subject of this work, we want reconstruct some f-essence models with the help of Chaplygin gas models. But before, for the pedagogical reason here we give some (may be) known facts on fermionic Chaplygin gas models for the usual Einstein-Dirac equation (see e.g. \cite{Saha10}). For the FRW metric, the  usual Einstein-Dirac equation has the form
	\begin{eqnarray}
	3H^2-\rho&=&0,\\ 
		2\dot{H}+3H^2+p&=&0,\\
				\dot{\psi}+1.5H\psi+i\gamma^0V_{\bar{\psi}}&=&0,\\ 
\dot{\bar{\psi}}+1.5H\bar{\psi}-iV_{\psi}\gamma^{0}&=&0,\\
	\dot{\rho}+3H(\rho+p)&=&0,
	\end{eqnarray} 
where    the kinetic term, the energy density  and  the pressure  take the form
  \begin{equation}
 Y=0.5i(\bar{\psi}\gamma^{0}\dot{\psi}-\dot{\bar{\psi}}\gamma^{0}\psi),\quad 
  \rho=V,\quad
p=Y-V.
  \end{equation}So  the usual Einstein-Dirac theory  we here understand as the particular f-essence with\begin{equation}
 K=Y-V.
  \end{equation}Our next step is the construction some Chaplygin gas models of dark energy for the model (4.7).  Consider examples.
 
 \subsection{The fermionic Chaplygin gas model} As first example, consider   the Chaplygin gas \cite{Kamenshchik}
   \begin{equation}
 p=-\frac{A}{\rho},
  \end{equation}
where $A$ is a positive constant. Here we note that the cosmological model based on the Chaplygin gas
was proposed for the first time in \cite{Kamenshchik}. Let $V=V(\bar{\psi}, \psi)=V(u)$.   Then, in terms of $a$, the system (4.1)-(4.5) has the following solution
  \begin{eqnarray}
	H&=&\pm 3^{-0.5}(A+Ba^{-6})^{0.25},\\ 
		\rho&=&(A+Ba^{-6})^{0.5},\\
				p&=&-A(A+Ba^{-6})^{-0.5},\\ 
				\psi_j&=&c_ja^{-1.5}e^{-iD},\quad j=1,2,\\
				\psi_l&=&c_la^{-1.5}e^{iD},\quad l=3,4,\\
					V&=&(A+Ba^{-6})^{0.5}.	
	\end{eqnarray} 
Here $B, c_j, c_l, c$, constants, $c=|c_1|^2+|c_2|^2-|c_3|^2-|c_4|^2$ and
	 \begin{equation}
D=\mp 0.5\sqrt{3}Bc^{-1}a^{-2}(A+Ba^{-6})^{-0.75}.
  \end{equation}
	Also we mention that  the scale factor is given by  \cite{Kamenshchik}
	 \begin{equation}
t=\frac{1}{6\sqrt[4]{A}}\left(\ln\frac{\sqrt[4]{A+Ba^{-6}}+\sqrt[4]{A}}{\sqrt[4]{A+Ba^{-6}}-\sqrt[4]{A}}-2\arctan\sqrt[4]{1+A^{-1}Ba^{-6}}\right).
  \end{equation}
  For the example, the EoS parameter is
   \begin{equation}
\omega=-1+\frac{B}{B+Aa^{6}}.
  \end{equation}
  Finally we present  expressions for the potential, the kinetic term and $u$:
  \begin{equation}
V=\sqrt{A+Bc^{-2}u^2}, \quad Y=Ba^{-6}(A+Ba^{-6})^{-0.5},
\quad u=ca^{-3}.
  \end{equation}
  We note that the relation between the Chaplygin gas and the Einstein-Dirac theory  was also explored in \cite{Saha10}. 
  \subsection{The fermionic generalized  Chaplygin gas model}
   The generalized  Chaplygin gas  has the following EoS \cite{Bento} 
   \begin{equation}
 p=-\frac{A}{\rho^\alpha},
  \end{equation}
where $A$ is a positive constant. For simplicity, we assume that  $V=V(\bar{\psi}, \psi)=V(u)$.   Then,  the system (4.1)-(4.5) has the following solution
  \begin{eqnarray}
	H&=&\pm 3^{-0.5}[A+Ba^{-3(\alpha+1)}]^{\frac{0.5}{\alpha+1}},\\ 
		\rho&=&[A+Ba^{-3(\alpha+1)}]^{\frac{1}{\alpha+1}},\\
				p&=&-A[A+Ba^{-3(\alpha+1)}]^{-\frac{\alpha}{\alpha+1}},\\ 
				\psi_j&=&c_ja^{-1.5}e^{-iD},\quad j=1,2,\\
				\psi_l&=&c_la^{-1.5}e^{iD},\quad l=3,4,\\
					V&=&[A+Ba^{-3(\alpha+1)}]^{\frac{1}{\alpha+1}}.	
	\end{eqnarray} 
Here $B, c_j, c_l, c$, constants, $c=|c_1|^2+|c_2|^2-|c_3|^2-|c_4|^2$ and
\begin{equation}
D=\mp \sqrt{3}Bc^{-1}\int a^{-3(1+\alpha)+3}\left[A+Ba^{-3(1+\alpha)}\right]^{-\frac{0.5+\alpha}{1+\alpha}}da.
  \end{equation}	
  For the example, the EoS parameter is
   \begin{equation}
\omega=-1+\frac{B}{B+Aa^{3(\alpha+1)}}.
  \end{equation}
  Finally we have the following expressions   for the potential, the kinetic term and $u$:
 \begin{eqnarray}
V&=&\left[A+Bc^{-(1+\alpha)}u^{1+\alpha}\right]^{\frac{1}{1+\alpha}}, \\ Y&=&Ba^{-3(1+\alpha)}\left[A+Ba^{-3(1+\alpha)}\right]^{-\frac{\alpha}{1+\alpha}},\\
u&=&ca^{-3}.
 \end{eqnarray} 
  \subsection{The fermionic modified Chaplygin gas model}
  The EoS of the modified Chaplygin gas  dark energy model is given by
   \cite{Benaoum}
    \begin{equation}
p=E\rho-\frac{A}{\rho^\alpha},
  \end{equation}
 where $A$ and $E$ are positive constants and $0\leq\alpha\leq 1.$  Using Eqs. (4.1)-(4.2), the  modified Chaplygin gas energy density evolves as \cite{Benaoum}
\begin{equation}
\rho=\left[A(1+E)^{-1}+Ba^{-3(1+\alpha)(1+E)}\right]^{\frac{1}{1+\alpha}},
  \end{equation}
 where $B$ is an integration constant.   As above, we assume that $V=V(\bar{\psi}, \psi)=V(u)$.   The corresponding solution of the  system (4.1)-(4.5) in terms of $a$  has the following form
  \begin{eqnarray}
	H&=&\pm 3^{-0.5}\left[A(1+E)^{-1}+Ba^{-3(1+\alpha)(1+E)}\right]^{\frac{0.5}{1+\alpha}},\\ 
		\rho&=&\left[A(1+E)^{-1}+Ba^{-3(1+\alpha)(1+E)}\right]^{\frac{1}{1+\alpha}},\\
				p&=&[EBa^{-3(1+\alpha)(1+E)}-A(1+E)^{-1}]\left[A(1+E)^{-1}+Ba^{-3(1+\alpha)(1+E)}\right]^{-\frac{\alpha}{1+\alpha}},\\ 
				\psi_j&=&c_ja^{-1.5}e^{-iD},\quad j=1,2,\\
				\psi_l&=&c_la^{-1.5}e^{iD},\quad l=3,4,\\
					V&=&\left[A(1+E)^{-1}+Ba^{-3(1+\alpha)(1+E)}\right]^{\frac{1}{1+\alpha}},	
	\end{eqnarray} 
where
	 \begin{equation}
D=\mp \sqrt{3}B(1+E)c^{-1}\int a^{-3(1+\alpha)(1+E)+3}\left[A(1+E)^{-1}+Ba^{-3(1+\alpha)(1+E)}\right]^{-\frac{0.5+\alpha}{1+\alpha}}da.
  \end{equation}
	
  In this case,  the EoS parameter is
   \begin{equation}
\omega=\frac{EBa^{-3(1+\alpha)(1+E)}-A(1+E)^{-1}}{A(1+E)^{-1}+Ba^{-3(1+\alpha)(1+E)}}.
  \end{equation}
  The corresponding  expressions for the potential, the kinetic term and $u$ read as:
 \begin{eqnarray}
V&=&\left[A(1+E)^{-1}+Bc^{-(1+\alpha)(1+E)}u^{(1+\alpha)(1+E)}\right]^{\frac{1}{1+\alpha}}, \\ Y&=&(1+E)Ba^{-3(1+\alpha)(1+E)}\left[A(1+E)^{-1}+Ba^{-3(1+\alpha)(1+E)}\right]^{-\frac{\alpha}{1+\alpha}},\\
u&=&ca^{-3}.
 \end{eqnarray} 
  \section{Reconstruction of f-essence. Fermionic   Chaplygin gas models of dark energy from f-essence}
  
  \subsection{The fermionic   Chaplygin gas model from f-essence}
Let us consider the following f-essence model
 \begin{equation}
  K= V_1\sqrt{1+V_2Y^2}, \end{equation}
  where $V_j=V_j(\bar{\psi}, \psi)$ are potentials so that we have $V_{jY}=0$. Then
  \begin{equation}
  p= V_1\sqrt{1+V_2Y^2},\quad \rho=-\frac{V_1}{\sqrt{1+V_2Y^2}}. \end{equation}
  Hence we get
  \begin{equation}
  p=-\frac{V_1^2}{ \rho}. \end{equation}
  It is the Chaplygin gas model. The intriguing characteristic of Chaplygin gas is that in the FRW metric, the energy density is
   \begin{equation}
  \rho=V_1\sqrt{1+CV_1^{-2}a^{-6}}, \quad C=const. \end{equation}
  Hence follows that, for example, at early times when $a<<CV_1^{-2})^{1/6}$, the gas behaves as a pressureless dust, $\rho =\sqrt{C}a^{-3}.$  Meanwhile it approaches asymptotically a "cosmological constant" $\rho=V_1$ at late times when $a>>(CV_1^{-2})^{1/6}$.  Note that for the f-essence (4.1) the equation of state parameter is \begin{equation}
  \omega=-1-V_2Y^2. \end{equation} 
  
  \subsection{The fermionic generalized Chaplygin gas model from f-essence}
  As is known that the Chaplygin gas model of dark energy suffers from strong observational pressure from explaining CMB anisotropies. This shortcoming is a alleviated in a  generalized Chaplygin gas model introduced.  Here we establish a connection between f-essence and the generalized Chaplygin gas model. To do it,  let us consider the following particular f-essence model
  \begin{equation}
  K= V_1(V_2Y^{\frac{\alpha+1}{\alpha}}+1)^{\frac{\alpha}{\alpha+1}}, \end{equation}
 where as above  $V_j=V_j(\bar{\psi}, \psi)$ are potential functions so that  $V_{jY}=0$. Then we have 
    \begin{equation}
  p=V_1(V_2Y^{\frac{\alpha+1}{\alpha}}+1)^{\frac{\alpha}{\alpha+1}},\quad \rho=-\frac{V_1}{(V_2Y^{\frac{\alpha+1}{\alpha}}+1)^{\frac{1}{\alpha+1}}}. \end{equation}
  Hence we get the generalized Chaplygin gas model
    \begin{equation}
  p=(-1)^{\alpha}\frac{V_1^{\alpha+1}}{\rho^\alpha}=-\frac{A}{\rho^\alpha}, \end{equation}
 where $A=(-V_1)^{\alpha+1}$. In the generalized Chaplygin gas model,  the energy density and  the equation of state parameter are given by \begin{equation}
  \omega=-1-V_2Y^{\frac{\alpha+1}{\alpha}} \end{equation} 
  and
   \begin{equation}
  \rho=[A+Ba^{-3(\alpha+1)}]^{\frac{1}{\alpha+1}},\quad B=const. \end{equation}
  Note that the original Chaplygin gas model (4.3) amounts to the case $\alpha=1$. Some properties of  the generalized Chaplygin gas model in non-flat universe were considered in\cite{Setare1}-\cite{Setare3}.
 
\section{Conclusion}
We reconstructed the  different particular models of f-essence in the homogeneous, isotropic and flat FRW universe.  In particular, the fermionic Chaplygin gas and  the fermionic generalized Chaplygin gas models of dark energy are presented. We also obtained the equation of state parameter of some selected models including for both the fermionic Chaplygin gas and  the fermionic generalized Chaplygin gas models. These results show that f-essence can describes the accelerated expansion of the universe.

\end{document}